\documentclass[aps,prl,twocolumn,showpacs]{revtex4}%
\usepackage{graphicx}
\usepackage{times,xspace}
\usepackage{amsbsy,amssymb,amsmath,bm}
\usepackage{amsmath}
\usepackage{amsfonts}
\usepackage{amssymb}%
\def\bbbc{{\mathchoice {\setbox0=\hbox{$\displaystyle\rm C$}\hbox{\hbox
to0pt{\kern0.4\wd0\vrule height0.9\ht0\hss}\box0}}
{\setbox0=\hbox{$\textstyle\rm C$}\hbox{\hbox
to0pt{\kern0.4\wd0\vrule height0.9\ht0\hss}\box0}}
{\setbox0=\hbox{$\scriptstyle\rm C$}\hbox{\hbox
to0pt{\kern0.4\wd0\vrule height0.9\ht0\hss}\box0}}
{\setbox0=\hbox{$\scriptscriptstyle\rm C$}\hbox{\hbox
to0pt{\kern0.4\wd0\vrule height0.9\ht0\hss}\box0}}}}

\begin{document}
\title[Specific Heat of SrCu$_{2}$(BO$_{3}$)$_{2}$]{High Field Specific Heat of 2D Quantum Spin System SrCu$_{2}$(BO$_{3}$)$_{2}$ }
\author{G.A. Jorge$^{1,}$$^{a}$}
\author{R. Stern$^{2}$}
\author{M. Jaime$^{1}$}
\author{N. Harrison$^{1}$}
\author{J. Bon\v{c}a$^{3,4}$}
\author{S.~El~Shawish$^{4}$}
\author{C.D. Batista$^{5}$}
\author{H.A. Dabkowska$^{6}$}
\author{B.D. Gaulin$^{6}$}

\affiliation{$^{1}$ MST-NHMFL, Los Alamos National Laboratory, Los  
Alamos, NM 87545,
USA.\linebreak\ $^{a}$ Also Departamento de F\'{\i}sica, Universidad de  
Buenos Aires, Bs. As., Argentina}

\affiliation{$^{2}$National Institute of Chemical Physics \& Biophysics  
(NICPB), Tallinn, Estonia.}

\affiliation{$^3$Department of Physics, FMF University of Ljubljana  
and $^4$J. Stefan Institute, Ljubljana, Slovenia}

\affiliation{$^5$ Theoretical Division,
Los Alamos National Laboratory, Los Alamos, NM 87545, USA}

\affiliation{$^{6}$Department of Physics and Astronomy, McMaster University, Hamilton, ON,
L8S 4C6, Canada}

\begin{abstract}
We report measurements of the specific heat of the quantum spin liquid system
SrCu$_{2}$(BO$_{3}$)$_{2}$ in continuous magnetic fields $H$ of up to 33~T.
The specific heat data, when combined with a finite temperature Lanczos
diagonalization of the Shastry-Sutherland Hamiltonian, indicates the presence
of a nearest neighbor Dzyaloshinsky-Moriya (DM) interaction that violates the
crystal symmetry for $H=0$. Moreover, the same DM interaction is required to
explain the observed ESR lines for $H \| c$. These results indicate that
spin-lattice coupling needs to be included in any realistic description of
this system.

\end{abstract}
\volumeyear{year}
\volumenumber{number}
\issuenumber{number}
\eid{identifier}
\date{January 19, 2004}
\received[Received text]{date}

\revised[Revised text]{date}

\accepted[Accepted text]{date}

\published[Published text]{date}

\maketitle




SrCu$_{2}$(BO$_{3}$)$_{2}$ is a quasi-two dimensional spin system with a
singlet dimer ground state \cite{smith91}. It is the only known realization of
the Shastry-Sutherland model \cite{shastry81}, and exhibits a sequence of
magnetization plateaux at high magnetic fields \cite{kageyama99,onizuka00}.
The unique behavior of this quantum spin liquid results from the interplay
between two different fascinating aspects of strongly correlated spin systems:
namely \textit{geometrical frustration} and \textit{strong quantum
fluctuations}. The spin $s=$~1/2 Cu$^{2+}$ ions that are responsible for the
magnetism are grouped in dimers within planes of the tetragonal SrCu$_{2}%
$(BO$_{3}$)$_{2}$ unit cell, with respective intra-dimer or nearest neighbor
($nn$) and inter-dimer or next nearest neighbor ($nnn$) separations of
2.9~\AA \hspace{0.05cm} and 5.1~\AA . The coupling constants are estimated to
be $J\sim$~80~K for $nn$ and $J^{\prime}\sim$~50~K for $nnn$~\cite{miyahara99}%
. The geometrical frustration of the spin-lattice leads to very localized
triplet excitations that have a tendency to crystallize at high magnetic
fields. This occurs when the concentration of triplets reaches certain values that are
commensurate with the underlying lattice, becoming
incompressible upon formation of a gapped structure. The magnetization
plateaux at $H_{p1}=27$~T, $H_{p2}=35$~T and $H_{p3}=42$~T are a direct
consequence of spin superstructures forming at triplet concentrations 1/8, 1/4
and 1/3 respectively.

Recent ESR experiments \cite{cepas01,cepas02,nojiri03} revealed the spin
triplet excitation energy to decrease linearly with increasing magnetic field,
extrapolating to zero at a field value ($H=22$~T) very close to H$_{p1}$.
However, on approaching H$_{p1}$, the experimental ESR data deviates from this
linear extrapolation, indicating a level anti-crossing between the first
triplet excitation and the ground state. The anti-crossing implies some mixing
between two states with different magnetization $M_{z}$ along the tetragonal
$c$-axis. This observation cannot be explained by the $U(1)$ invariant models
(which are symmetric under rotations around the $c$-axis) proposed in previous
works, for which $M_{z}$ is a good quantum number.

In this letter we argue that an intra-dimer Dzyaloshinsky-Moriya (DM)
interaction, which violates the observed crystal symmetry at $100K \leq T \leq 395K$
\cite{sparta01}, is required to explain the low temperature specific heat of
SrCu$_{2}$(BO$_{3}$)$_{2}$ in magnetic fields $H>$~18~T. We also show that
this interaction gives rise to the ESR transitions between the ground state
and the single triplet excitations that are observed for $H \| c$
\cite{cepas01,cepas02,nojiri03}. These results suggest that there is a structural
phase transition at low temepratures that lowers the crystal symmetry. Such a transition
could be driven by a strong spin-lattice interaction  that would be a relevant ingredient 
to explain the magnetization plateaux which are observed in this system.

The single crystal sample of SrCu$_{2}$(BO$_{3}$)$_{2}$ used in this study was
grown by the floating zone technique. Stoichiometric amounts of CuO,
SrCO$_{3}$ and B$_{2}$O$_{3}$ were mixed, pre-annealed, and then annealed at
870 ${{}^{o}}$C. Finally, the powder was regrinded, pelletized and annealed in
O$_{2}$ several times. Rods were formed by hydrostatic pressing and the growth
was performed in a Crystal System Optical Furnace at a growth speed of
0.25mm/h in O$_{2}$. No additional flux was applied \cite{dabkowska03}. The
measurements of the specific heat $C(T,H)$ of SrCu$_{2}$(BO$_{3}$)$_{2}$ in
continuous magnetic fields up to 33~T were performed on two oriented single
crystal pieces of 12.34~mg and 13.92~mg. Both were measured with $H$ applied
along the tetragonal $c$-axis and within the $ab$ planes. A calorimeter made
of plastic materials and silicon was used, employing a thermal relaxation time
technique optimized for rapid data acquisition~\cite{jaime00,jorge}. The
magnetization $M_{z}$ of a piece of sample of approximate dimensions 1.5
$\times$\ 0.9 $\times$\ 0.5 mm$^{3}$ was measured as a function of field using
a sample-extraction magnetometer in a 400~ms, 45~T pulsed magnet
provided by the National High Magnetic Field Laboratory at Los
Alamos~\cite{jorge03}. The small size of the sample, placed in good thermal
contact with liquid $^{3}$He or $^{4}$He below $T=4$~K, combined with the
relatively slow field sweep of the magnet helped minimize magnetocaloric
effects so as to achieve an isothermal experiment~\cite{jaime02}. For
characterization purposes, supplementary $M_{z}$ versus temperature $T$
measurements and specific heat measurements were made at lower fields using a
commercial Quantum Design\texttrademark~MPMS (SQUID magnetometer). Meanwhile,
numerical simulations of the Shastry-Sutherland model, with which the
experimental data are compared, were performed on a 20-site square lattice
using the finite temperature Lanczos (FTL)
method~\cite{jacklic96,jprev,bonca02}.

To describe the present system, we consider the following Heisenberg
Hamiltonian on a Shastry-Sutherland lattice \cite{shastry81}:
\begin{align}
H_{s}  &  = J \sum_{\langle\mathbf{i, j} \rangle} \mathbf{S_{i} \cdot S_{j}} +
J^{\prime}\sum_{\langle\mathbf{i, j} \rangle^{\prime}} \mathbf{S_{i} \cdot
S_{j}}\nonumber\\
&  +\sum_{\langle\mathbf{i \rightarrow j} \rangle} \mathbf{D} \cdot
(\mathbf{S_{i} \times S_{j}}) + \sum_{\langle\mathbf{i \rightarrow j}
\rangle^{\prime}} \mathbf{D^{\prime}} \cdot(\mathbf{S_{i} \times S_{j}}).
\label{Hamil}%
\end{align}
Here, $\langle{i,j} \rangle$ and $\langle{i,j} \rangle^{\prime}$ indicate that
$\mathbf{i}$ and $\mathbf{j}$ are $nn$ and $nnn$ respectively. The Hamiltonian
includes $nn$ ($\mathbf{D}$) and $nnn$ ($\mathbf{D^{\prime}}$) DM
interactions. The arrows indicates that the corresponding bonds have a
particular orientation. The quantization axis ${\hat{\mathbf{z}}}$ is parallel
to the $c$-axis. The $nnn$ DM interaction has already been considered in
previous papers \cite{miyahara03} to explain the position of the single
triplet excitations observed with ESR \cite{cepas01,nojiri03}, far infra-red
\cite{room00} and inelastic neutron scattering measurements \cite{kageyama00}.
From the splitting between the two single-triplet excitations at
$\mathbf{q=0}$, the $\mathbf{D^{\prime}}=$ is estimated to be:
$\mathbf{D^{\prime}}=$~2.1~K${\hat{\mathbf{z}}}$. The orientation of the $nnn$
bonds is given in Ref.~\cite{miyahara03}. According to the crystal symmetry
\cite{sparta01}, the $xy$ component of $\mathbf{D}$ is perpendicular to the
corresponding dimer. However, as is explained below, a non-zero $z$ component
of $\mathbf{D}$, which is not allowed by the observed crystal symmetry at
high temperatures ($100K \leq T \leq 395K$), is required to explain the specific heat 
and the ESR data a a function of the applied  field $H$.

\begin{figure}
\includegraphics[scale=0.3]{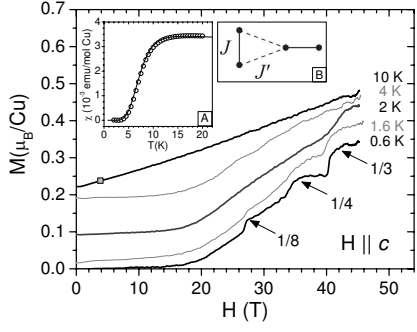}
\caption{Magnetization vs field for SrCu$_{2}$(BO$_{3}$)$_{2}$ at different
temperatures between 0.6~K and 10~K, as indicated, reveal a gradual evolution
of the magnetization plateaux. The square point on the 10~K curve was used to
compare with SQUID magnetometer data in order to obtain the magnetization
units. Inset A: Magnetic susceptibility measured at H = 4~T in a SQUID
magnetometer (circles). The solid line is the susceptibility calculated with
the FTL method. Inset B: Two copper dimers in the CuBO$_{3}$ plane where the
coupling constants $J$ (nn) and $J^{\prime}$ (nnn) are indicated.}%
\label{fig1}%
\end{figure}

Figure \ref{fig1} shows the magnetization $M_{z}(H)$ measured as a function of
magnetic field, in units of $\mu_{B}$/Cu determined upon cross-calibration
with SQUID magnetometry data. In addition to the plateaux already mentioned,
there is a small excess contribution to $M_{z}$ identified in our data over
the entire field range. This additional source of magnetization has been
observed before in SQUID magnetometry data \cite{kageyama99}, although for
reasons which are unknown to us, it is absent in published pulsed magnetic
field data. For our samples, both SQUID magnetometry data and low field pulsed
magnetic field data evidence a finite excess susceptibility of approximately
$0.115 \times10^{-3}$ emu/mol~Cu, probably due to crystalline defects. Good
agreement with the expected magnetization values at the plateaux is obtained
by subtracting this value. Better agreement is obtained by subtracting a
scaled Brillouin function with an initial slope of $0.14 \times10^{-3}$
emu/mol~Cu and a characteristic temperature of 5~K (see Fig.~\ref{fig1}).
After either of these substractions, there remains a finite value of $M_{z}$
at very low temperatures and magnetic fields that increases linearly with $H$.

In the inset of Fig.~\ref{fig1} we compare the measured magnetic
susceptibility $\chi(T)$ (after subtracting a small constant value of $0.14
\times10^{-3}$ emu/mol~Cu) and the curve obtained with the FTL method that is
described below. We get an excellent agreement for: $J=74$~K, $J^{\prime
}=0.62J$, \textbf{D}=(2.2K,$\pm$ 2.2K,5.2K) (the sign is different for each
dimer in the unit cell), \textbf{D}'=(0,0,2.2K). The values of the
$g$-factors, $g_{\parallel}=2.15$ and $g_{\perp}=2.08$, have been obtained
from a comparison between our theoretical calculations \cite{janez04} (see
Fig.~\ref{fig4}) and the ESR spectra \cite{cepas01,nojiri03}.

\begin{figure}
\includegraphics[scale=0.3]{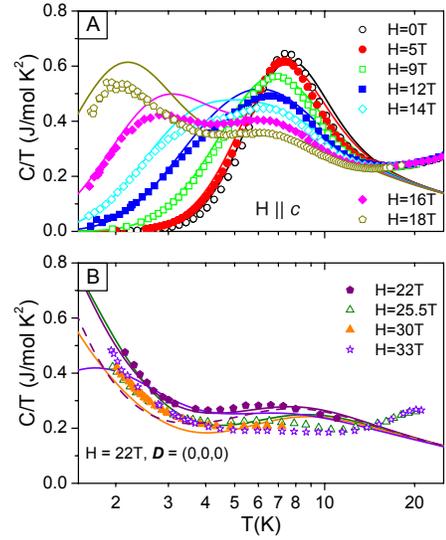}
\caption{ Measured specific heat divided by $T$ (symbols) vs. $T$ compared
with the calculated one for the Hamiltonian of Eq.~\ref{Hamil} (solid lines)
for the magnetic field along the $c$-axis: A) $0\leq H \leq18$~T and B)
$20\leq H \leq33$~T. The parameters are the same ones used to compute the
magnetic susceptibility. The dashed line is the calculated $C/T$ for
$\mathbf{D}=0$ and $H=22$T.}%
\label{fig2}%
\end{figure}

\begin{figure}
\includegraphics[scale=0.3]{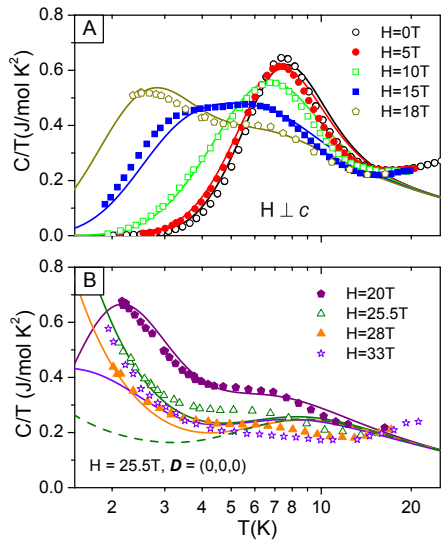}
\caption{ Measured specific heat divided by $T$ (symbols) vs. $T$ compared
with the calculated one for the Hamiltonian of Eq.~\ref{Hamil} (solid lines)
for the magnetic field perpendicular to the $c$-axis:A) $0\leq H \leq18$~T and
B) $22\leq H \leq33$~T. The parameters are the same ones used to compute the
magnetic susceptibility. The dashed line is the calculated $C/T$ for
$\mathbf{D}=0$ and $H=25.5$T.}%
\label{fig3}%
\end{figure}

\begin{figure}
\includegraphics[scale=0.3]{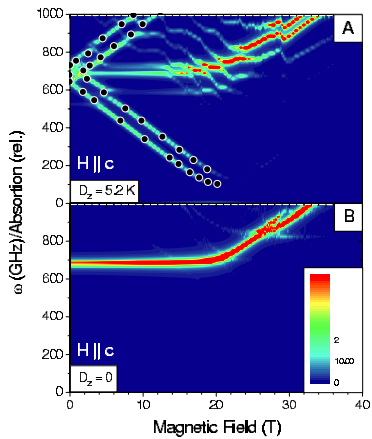}
\caption{ Contour color plot for the ESR spectrum for $H\Vert c$ calculated
with the Lanczos method in a 20 sites cluster for a) $D_{z}=5.2$~K and b)
$D_{z}=0$. The values of the other parameters are the same ones used to
compute the magnetic susceptibility. The experimental data points in a) are
from Cepas et al., [6].}%
\label{fig4}%
\end{figure}

In Fig.~\ref{fig2}, we show the specific heat divided by temperature,
$C(T,H)/T$, for different values of the magnetic field applied along the
$c$-axis. The primary feature in the low temperature specific heat is a broad
anomaly centered at $T=8.5$~K that is gradually depressed by increasing $H$.
This anomaly has been attributed \cite{hofmann02} to $S_{z}=0$ dimer
excitations. Here, however, we observe a small shift in temperature as
function of $H$, indicating the involvement of states with $S_{z}\neq$~0. For
$H\geq$~12~T, a second anomaly develops at lower temperatures, which we
attribute to single-triplet excitations situated 3~meV above the ground state
in zero field. The Zeeman interaction causes this triplet states to move to
lower energies with increasing $H$. Figure~\ref{fig3} shows similar results
for $H \perp c$.

In Figs.~\ref{fig2} and \ref{fig3}, we also compare the experimental results
with the results of a numerical simulation of $C(T,H)/T$ made using the FTL
method \cite{jacklic96,jprev}. This method is based on the Lanczos procedure
of exact diagonalization, and uses a random sampling over initial wave
functions specially adapted for calculation of thermodynamic properties. All
the results were computed on a tilted square lattice of $N=20$ sites. There
are many advantages of this method over the conventional Quantum Monte Carlo
(QMC) methods, which are as follows: first, the minus-sign problem that
usually appears in QMC calculations of frustrated spin systems is absent;
second, the method connects the high- and low-temperature regimes in a
continuous fashion, enabling the entropy density and specific heat (per unit
cell) to be computed as expectation values (\textit{i.e.} $s=k_{B} ln
Z/N+\langle H\rangle/NT$, where $Z$ is the statistical sum). The specific heat
is then given by $C_{V}=T(\partial s/\partial T)=k_{B}(\langle H^{2}%
\rangle-\langle H\rangle^{2})/NT^{2}$. The main limitation to the validity of
the results originates from finite-size effects which occur when $T<T_{fs}$.
The actual value of $T_{fs}$ depends strongly on the particular physical
properties of the system. For gapless systems, $T_{fs}$ can be defined by way
of the thermodynamic sum $\bar Z(T)=\mathrm{Tr~exp}(-(H-E_{0})/T)$, on
condition that $\bar Z(T_{fs}) = Z^{*}\gg1$ \cite{jprev}. In the present case,
this condition can be relaxed ($Z^{*}> 1$) owing to the existence of a gap in
the excitation spectrum when $J^{\prime}/J< 0.7$ and to the almost localized
nature of the lowest excited states - triplet excitations. By comparing
results obtained on two different systems with $N=16$ and $N=20$ sites, we
estimate $T_{fs}<1$~K.

For $H<18$~T, the agreement between theory and experiment is very good,
regardless of the inclusion of the $nn$ DM interaction. Finite size effects
are also very small, for $H<18$~T, due to the localized nature of the
single-triplet excitations \cite{kageyama00}. When $H$ approaches 22~T for
$H\| c$ (or 25~T for $H\perp c$), however, the inclusion of this interaction
is required to explain the measured $C(T,H)/T$ at low temperatures. For $H\|
c$, this is explained by the fact that $\mathbf{D}$ is the only interaction
that violates the conservation of $M_{z}$, by mixing the $M_{z}=0$ singlet
ground state of $H(\mathbf{D}=0)$ with the single-triplet excited state with
$M_{z}=\pm1$. This mixing becomes effective only when the energy difference
between both levels is comparable to $|\mathbf{D}|$. For $H\perp c$, the same
type of mixing is produced by the $z$-component of $\mathbf{D}$. In other
words, the level crossing that would occur in absence of the DM interactions
is replaced by level anti-crossing. This can be seen in Figs.~\ref{fig2}b and
\ref{fig3}b where we also show the calculated $C/T$ for $\mathbf{D}=0$ and
$H=22T$ ($H=25.5T$) for $H \| c$ ($H \perp c$). In absence of the $DM$
interactions, the level crossing generates a peak of $C/T$ at $T=0$ which is
not consistent with the experiment. In contrast, the level anti-crossing moves
this peak to higher temperatures in agreement with the experimental data. The
anti-crossing occurs for different values of $H$ in the different field
orientations due to the anisotropy of the $g$-factor and the fact that
$\mathbf{D}$ is parallel to the $c$ axis \cite{cepas01,nojiri03}. At high
temperatures ($T > 20$~K), the specific heat data deviates from the
theoretical prediction owing to significant phonon contributions.

For $T\sim10$~K and fields $H>20$~T, there are small deviations between the
experimental curves and the calculations, which can be attributed to the inter
planar antiferromagnetic interaction $J^{\prime\prime}/J \sim0.21$ that
becomes relevant when the concentration of triplet excitations increases.

The components of the $nn$ DM interaction are constrained by the crystal
symmetry at low temperatures \cite{sparta01}. According to this symmetry, the
$z$-component of $\mathbf{D}$ must be zero. Although a non-zero DM vector of
the form $(d,\pm d,0)$ improves considerably the agreement between experiment
and theory for $H \| c$, it does not reproduce our experimental data in the
proximity of $H_{p1}$= 27 T for $H \perp c$. Ultrasonic experiments
\cite{zherlitsyn} indicate that in this region the lattice gets strongly
distorted by the application of the magnetic field . We speculate that this
lattice distortion increases the $z$-component of the
DM vector to lower the magnetic energy (it increases the level anti-crossing).

This speculation is further supported by the measured ESR spectrum for $H\Vert
c$ \cite{cepas01,nojiri03}. We show the ESR spectrum as a function of $H$ for
$D_{z}=5.2$ (Fig.~\ref{fig4}a) and $D_{z}=0$K (Fig.~\ref{fig4}b) calculated
with the Lanczos method \cite{janez04}. More specifically, we are computing the
dynamical susceptibility along the direction  perpendicular to the 
applied field using the method introduced in Ref.~\cite{gagliano}.
As it is pointed out in Ref.~\cite{cepas02}, the
observed ESR trasitions between the ground state and the single-triplet
excitations are not allowed by the observed crystal symmetry at $H=0$
\cite{sparta01}. In Fig.~\ref{fig4}a we show that these ESR transitions can be
explained with a non-zero value of $D_{z}$, while the corresponding ESR lines
are not present if $D_{z}=0$ (Fig.~\ref{fig4}b). None of the other components
of $\mathbf{D}$ or $\mathbf{D^{\prime}}$ can reproduce these ESR lines. Based
on these observations, we propose that the crystal symmetry is lowered at low
temepratures due to a strong spin-lattice interaction. Since
the lattice distortion depends on the applied field \cite{zherlitsyn}, we
expect $D_{z}$ to be an increasing function of $H$ (although we used a
constant value $D_{z}=5.2$K for our calculation). 

With the exception of $\mathbf{D}$, all other physical parameters used in the
model were determined from previous experiments. The values of $g_{\parallel}$
and $g_{\perp}$ are obtained from the ESR spectra\cite{cepas01,nojiri03}. By
including $\mathbf{D}$, however, we are able to account simultaneously for the
ESR spectra as a function of the applied  field (see also Ref. \cite{janez04}), 
the temperature dependence of the susceptibility, and the low temperature specific heat data
for $H \gtrsim18$~T.

In summary, we have measured the specific heat as a function of temperature in
continuous magnetic fields up to 33~T. An excellent fit to the $C(T,H)/T$ data
for both field orientations is obtained for a $nn$ exchange constant $J=74$~K,
a ratio $J^{\prime}/J=0.62$, a $nn$ Dzyaloshinsky-Moriya interaction constant
$|\mathbf{D}|=6.1$~K, and a $nnn$ Dzyaloshinsky-Moriya interaction constant
$|\mathbf{D}^{\prime}|=2.2$~K. A non-zero value of $D_{z}$, that is not
allowed by the observed crystal symmetry at $H=0$, is required to explain both
the specific heat data for $H \gtrsim18$~T and the observed ESR
\cite{cepas01,nojiri03} spectrum for $H \| c$. This suggests that a lattice
distortion that lowers the crystal symmetry is induced at low temperatures. 
A more detailed comparison between the calculated ESR spectrum and the
experiment will be presented elsewhere \cite{janez04}.

This work was sponsored by the US DOE under contract W-7405-ENG-36. R.S. was
supported by the National High Magnetic Field Laboratory and Estonian Science
Foundation grant No.4931. J.B. acknowledges the financial support of Slovene
MESS. Work performed at the National High Magnetic Field Laboratory is
supported by the National Science Foundation(DMR90-16241), the Department of
Energy and the State of Florida.


\end{document}